\documentstyle[twocolumn,aps,epsf]{revtex}
\begin{document}
\draft
\title{Quantum correlations are not contained in the initial
state\thanks{Accepted for publication in Phys. Rev. A.}}
\author{Ad\'{a}n Cabello\thanks{Electronic address:
fite1z1@sis.ucm.es}}
\address{Departamento de F\'{\i}sica Aplicada,
Universidad de Sevilla, 41012 Sevilla, Spain}
\date{\today}
\maketitle
\begin{abstract}
Two proofs are presented which show that quantum mechanics
is  incompatible with the following assumption:
all possible correlations
between subsystems of an individual isolated composite quantum
system are contained in the initial quantum state of the whole system,
although just a subset of them is revealed by the actual experiment.
\end{abstract}

\narrowtext

\section{Introduction}
Since the seminal work of Einstein, Podolsky, and Rosen
\cite{EPR35}, the entanglement between quantum
variables pertaining two different parts of
a composite system has been considered {\em the} characteristic trait
of quantum mechanics \cite{Schrodinger35}.
Recently discovered phenomena involving composite systems of more than two
parts, such as quantum dense coding \cite{qdc1,qdc2},
teleportation of quantum states \cite{telep1,telep2,telep3}, and
entanglement swapping \cite{telep1,es2,es3,es4},
have in common that they exploit the fact that in a multiparticle system,
entanglement between two
parts can be considered itself an entangled property \cite{KZ}.
The implications of these phenomena to
several proposed interpretations of quantum mechanics are currently subject
of investigation \cite{Vaidman}.
In this context, I have argued recently that a certain
interpretation of quantum mechanics based on the assumption that
correlations between subsystems of an individual isolated composite quantum
system are real objective local properties of that system
\cite{Mermina,Merminb,Merminc} is inconsistent \cite{Adan}.
My argument was based on an example: consider
two pairs of spin-$\frac{1}{2}$
particles prepared so that both pairs are in the singlet state
(the first pair is
composed by particles 1 and 2, and the second by particles 3 and 4).
By performing one among two alternative measurements on particles 2 and 3,
one can choose between two types of
correlations for particles 1 and 4 (spacelike separated from the
measurement on particles 2 and 3): they can be in a pure factorizable
state or, alternatively, in a pure maximally entangled state \cite{Adan}.
This then would allow one
to choose nonlocally the type of correlations between two distant parts.
In my opinion, this is inconsistent with the assumption that
such correlations were local objective properties.
However, in \cite{Adan} I wrote `I do not mean
that the internal correlations
between particles 1 and 4 ``change'' after a spacelike separated
experiment (this does not happen in the sense that no new internal
correlations are ``created'' that were not
``present'' in the reduced density
matrix for the system 1 and 4 before
any interaction), but that the type of
internal correlations (and therefore,
\ldots {\em the reality}) of an individual
isolated system can be chosen at distance.'
So, implicitly, I admitted that all such possible correlations
between two parts were present somehow in the initial quantum state
of the whole system, although just a subset of them is revealed
by the actual experiment. The aim of this note is to show that
even such an innocuous-looking assumption is incompatible
with quantum mechanics.
For that purpose I will present two proofs in which such
an assumption leads to a contradiction. The first is a
Greenberger-Horne-Zeilinger-like proof \cite{GHZ1,GHZ2,GHZ3}
involving three {\em pairs} of
spin-$\frac{1}{2}$ particles. This proof does not require inequalities
nor probabilities, indeed it also admits a reading
as a multiplicative proof of the Kochen-Specker theorem \cite{KS1,KS2}
in a Hilbert space of dimension sixty four. The second proof
is even simpler. It is a Hardy-like
proof without inequalities (but with probabilities) \cite{Hardy}
involving two pairs of spin-$\frac{1}{2}$ particles.

\section{GHZ-like proof of impossibility of preexistent correlations}
For the first proof consider three pairs of spin-$\frac{1}{2}$ particles
labeled from 1 to 6. The Hilbert space in which we
describe the spin state of this system has dimension sixty four.
I will call it ${\cal H}_{64}$.
Let $A_{ij}$ be the non-degenerate operator acting on the
four-dimensional subspace
of particles $i$ and $j$, defined as
\begin{equation}
A_{ij}=2\,\hat \alpha _{ij}^{++}+\hat \alpha _{ij}^{+-}-
\hat \alpha _{ij}^{-+}-2\,\hat \alpha _{ij}^{--},
\label{factoperator}
\end{equation}
where $\hat \alpha _{ij}^{+-}$ is the projection operator
onto the state
$\left| {\alpha^{+-} } \right\rangle _{ij}=\left| +
\right\rangle _i\otimes \left| - \right\rangle _j$, etc.
Let $B_{ij}$ be the non-degenerate
Bell operator \cite{BMR92} defined as
\begin{equation}
B_{ij}=2\,\hat \phi _{ij}^++\hat \psi _{ij}^+-
\hat \psi _{ij}^--2\,\hat \phi _{ij}^-,
\label{Belloperator}
\end{equation}
where $\hat \phi _{ij}^+$ is the projection operator
onto the state $\left| {\phi ^+ } \right\rangle _{ij}$, etc.,
being
\begin{eqnarray}
\left| {\phi ^\pm } \right\rangle _{ij}=
{\scriptstyle {1 \over {\sqrt 2}}}
\left( {\left| + \right\rangle _i\otimes \left| +
\right\rangle _j\pm
\left| - \right\rangle _i\otimes \left| -
\right\rangle _j} \right), \\
\left| {\psi ^\pm } \right\rangle _{ij}=
{\scriptstyle {1 \over {\sqrt 2}}}
\left( {\left| + \right\rangle _i\otimes \left| -
\right\rangle _j\pm
\left| - \right\rangle _i\otimes \left| +
\right\rangle _j} \right),
\end{eqnarray}
the four Bell states \cite{BMR92},
which form an orthogonal basis for the corresponding
four-dimensional subspace.

Consider now the four operators acting on ${\cal H}_{64}$ defined as
\begin{eqnarray}
A_{12}A_{34}B_{56}=A_{12}\otimes A_{34}\otimes B_{56},
\label{op1}\\
A_{12}B_{34}A_{56}=A_{12}\otimes B_{34}\otimes A_{56},
\label{op2}\\
B_{12}A_{34}A_{56}=B_{12}\otimes A_{34}\otimes A_{56},
\label{op3}\\
B_{12}B_{34}B_{56}=B_{12}\otimes B_{34}\otimes B_{56}.
\label{op4}
\end{eqnarray}
As it can be easily checked, any of these four operators has eigenvalues
$\pm2^k$, with $k=0,1,2,3$. In addition,
the four operators are mutually commutative so they possess a set
of common eigenvectors.
Let one of these common eigenvectors
be the initial state of the six-particle system;
for instance, the state
$\left| {\mu } \right\rangle$ defined
by the following eigenvalue equations:
\begin{eqnarray}
A_{12}A_{34}B_{56}\left| {\mu } \right\rangle & = &
\left| {\mu } \right\rangle,
\label{mu1} \\
A_{12}B_{34}A_{56}\left| {\mu } \right\rangle & = &
\left| {\mu } \right\rangle,
\label{mu2} \\
B_{12}A_{34}A_{56}\left| {\mu } \right\rangle & = &
\left| {\mu } \right\rangle,
\label{mu3} \\
B_{12}B_{34}B_{56}\left| {\mu } \right\rangle & = &
- \left| {\mu } \right\rangle.
\label{mu4}
\end{eqnarray}
Note that the four respective eigenvalues
($1,1,1$, and $-1$, in this case)
are not independent since they must
obey the same functional relations
satisfied by the four operators. In particular,
since the product of the four operators
is a {\em negative} operator (i.e., all
its eigenvalues are negative numbers) with eigenvalues $-16^m$,
with $m=0,1,2,3$, then the product of their four eigenvalues
must be negative.

Now let us assume, as it is in \cite{Mermina,Merminb,Merminc}, that
all the correlations between subsystems of the composed system
are real objective internal local properties of such subsystems. In
particular, consider three subsystems: the first is
composed by particles 1 and 2, the second by particles 3 and 4,
and the third by particles 5 and 6.
We will assume that all possible correlations
between particles 1 and 2 (for instance) are encoded
in the initial state for the whole system,
and they do not depend on any interaction
experienced by the other subsystems, so they cannot change
(in particular, they cannot be created)
as a result of any experiment performed on particles 3 to 6
(supposed to be spacelike separated from particles 1 and 2).

Now consider three observers, each having access to one pair of
particles. On each pair, they may measure either $A_{ij}$ or
$B_{ij}$, without disturbing the other pairs. The results
of these measurements will be called $a_{ij}$ or $b_{ij}$, respectively.
Since these results must satisfy the same functional relations
satisfied by the corresponding operator, then, from
(\ref{mu1}), we can predict that, if $A_{12}$, $A_{34}$, and, $B_{56}$
are measured, the results satisfy
\begin{equation}
a_{12}a_{34}b_{56} = 1.
\label{nu1}
\end{equation}
Analogously, from Eqs. (\ref{mu2})-(\ref{mu4}),
the results of other possible measurements satisfy
\begin{eqnarray}
a_{12}b_{34}a_{56} & = & 1,
\label{nu2} \\
b_{12}a_{34}a_{56} & = & 1,
\label{nu3} \\
b_{12}b_{34}b_{56} & = & -1.
\label{nu4}
\end{eqnarray}
We can associate each one of the eigenvalues $a_{ij}$ and $b_{ij}$
with a type of correlation between particles $i$ and $j$
initially hidden in the original state of the system, but
``revealed'' by performing measurements on the two {\em other}
distant pairs. For example, if $B_{12}$ and $B_{34}$ are measured
and their results are both $1$, then one can predict with certainty
that particles 5 and 6 are in the {\em singlet} state, and
since arriving to this conclusion does not
require any real interaction on particles 5 and 6,
then we assume that the spins of particles 5 and 6 were initially
correlated in the singlet state (i.\ e., the same spin component
of particles 5 and 6 would have opposite signs), so
we assign the value $-1$ for the observable $B_{56}$
to the initial state $\left| {\mu } \right\rangle$.
Alternatively, since a different
measurement on particles 1 to 4 (for instance,
by measuring $A_{12}$ instead of $B_{12}$) allows one to predict
with certainty,
without interacting with particles 5 and 6, how the $z$ spin components
of particles 5 and 6 are correlated, and since this
information do not require any real interaction on particles 5 and 6,
then we suppose that it was encoded somehow in the initial state
of the whole system (so we assign to the initial state one
of the eigenvalues of $A_{56}$). Such predictions with certainty
and without interaction would lead us to assign values
to the six types
of correlations given by $A_{12}$, $B_{12}$, $A_{34}$, $B_{34}$,
$A_{56}$, and $B_{56}$.
However, such an assignment cannot be consistent with the rules of
quantum mechanics because
the four equations (\ref{nu1})-(\ref{nu4}) cannot
be satisfied simultaneously, since the product of their left-hand sides is
a positive number (because each value appears twice), while
the product of the right-hand sides is $-1$.
Therefore, the whole information
on the correlations between the particles of the three pairs
cannot be encoded in the initial state as we assumed.

\section{Proof of the Kochen-Specker theorem in ${\cal H}_{64}$}
A similar argument could be developed starting from any common
eigenvector of the four operators (\ref{op1})-(\ref{op4}).
In fact, including these four operators, the argument
can be rearranged as a state-independent
proof of the Kochen-Specker theorem
\cite{KS1} (like the one proposed in \cite{KS2})
in a sixty four dimensional Hilbert space.
This proof is summarized in Fig. 1.


Fig. 1 contains ten operators: the four operators (\ref{op1})-(\ref{op4})
acting on the whole system, and six operators acting only on pairs
of particles (two operators for each pair). The four operators on
each of the five lines are mutually commutative. As told before,
the product of the four operators on the horizontal line is
a {\em negative} operator, and as can be easily verified, the product
of the four operators on each of the other lines is one (and the
same) {\em positive} operator (with eigenvalues $4^n$, with $n=0,1,2,3$).
It is easily checked that it is impossible
to ascribe one of their eigenvalues to the ten operators,
satisfying all the same functional relations that are satisfied by
the corresponding operators.

\section{Hardy-like proof of impossibility of preexistent correlations}
The second argument against the possibility
of predefined correlations is simpler. It requires just
two pairs of spin-$\frac{1}{2}$ particles.
Consider the following initial state:
\begin{eqnarray}
\left| \eta  \right\rangle =
{\scriptstyle{1 \over {2\sqrt 3}}} (  \left| {+-+-} \right\rangle -
\left| {+--+} \right\rangle  \nonumber \\
 \mbox{} - \left| {-++-} \right\rangle -3 \left| {-+-+} \right\rangle ),
\label{lucien}
\end{eqnarray}
where $\left| {+-+-} \right\rangle=\left| {+} \right\rangle_{1} \otimes
\left| {-} \right\rangle_{2} \otimes \left| {+} \right\rangle_{3} \otimes
\left| {-} \right\rangle_{4}$, etc.
As it can be easily checked, this state has the following four properties:
\begin{eqnarray}
P_\eta \left( {\left. {\psi _{12}^-}\, \right|\alpha _{34}^{+-}} \right)
& = & 1,
\label{p1} \\
P_\eta \left( {\left. {\psi _{34}^-}\, \right|\alpha _{12}^{+-}} \right)
& = & 1,
\label{p2} \\
P_\eta \left( {\alpha _{12}^{+-},\alpha _{34}^{+-}} \right)
& = & {\scriptstyle{1 \over {12}}},
\label{p3} \\
P_\eta \left( {\psi _{12}^-,\psi _{34}^-} \right)
& = & 0.
\label{p4}
\end{eqnarray}
Property (\ref{p1}) tells us that on every copy of the system
initially prepared in the state (\ref{lucien}) in
which the result of measuring $A_{34}$ is $\alpha _{34}^{+-}$, one can
predict (with certainty and without interacting with them) that particles
1 and 2 are in the singlet state. Therefore, we conclude
that in that subensemble of copies, particles 1 and 2
were initially in the singlet state.
Analogously, property
(\ref{p2}) tells us that on every copy of the system
initially prepared in the state (\ref{lucien}) in
which the result of measuring $A_{12}$ is $\alpha _{12}^{+-}$, one can
predict (with certainty and without interacting with them) that
particles 3 and 4 are in the singlet state. Therefore, we conclude
that in that subensemble, particles 3 and 4
were initially in the singlet state.
Property (\ref{p3}) reveals that the intersection between
the two subensembles defined before is not zero, since
there is a non-zero probability to obtain simultaneously
the two conditions defining such subensembles.
Assuming that
the predicted correlations were encoded in the
initial state of the system, the previous properties
would lead us to conclude that the probability of
finding both pairs of particles in the singlet state
is greater or equal than the probability of finding
simultaneously the corresponding conditions, given in
(\ref{p3}). However, property (\ref{p4}) shows
that this is no so. In fact, the probability
of finding two singlets is zero. Therefore, the
assumption that these correlations were contained
in the initial state is untenable.

\section{Conclusions}
``No-go'' proofs show that in QM local observables cannot have
predefined values \cite{GHZ1,GHZ2,GHZ3,KS2,Hardy}.
In this paper I have shown how, by duplicating the number of involved particles,
these proofs can be rearranged so as to exclude the possibility of predefined local
correlations between two particles of a composite system. This impossibility would
be taken into account in any attempt to describe phenomena
such as quantum dense coding, teleportation of quantum states or
entanglement swapping in a consistent interpretation of QM. In particular, this
impossibility of preexistent correlations constraints any further development
of the tentative interpretation proposed in \cite{Mermina,Merminb,Merminc}.\\

\section*{Aknowledgments}
The author thanks
Gonzalo Garc\'{\i}a de Polavieja for fruitful discussions,
and Christopher Fuchs, David Mermin, and Asher Peres for useful feedback.
Also, thanks are due to Jos\'{e} Luis Cereceda for pointing out a mistake in a
previous version.
This work was financially supported by the Universidad de Sevilla (OGICYT-191-97)
and the Junta de Andaluc\'{\i}a (FQM-239).


\begin{figure}
\epsfxsize=6.25cm
\epsffile{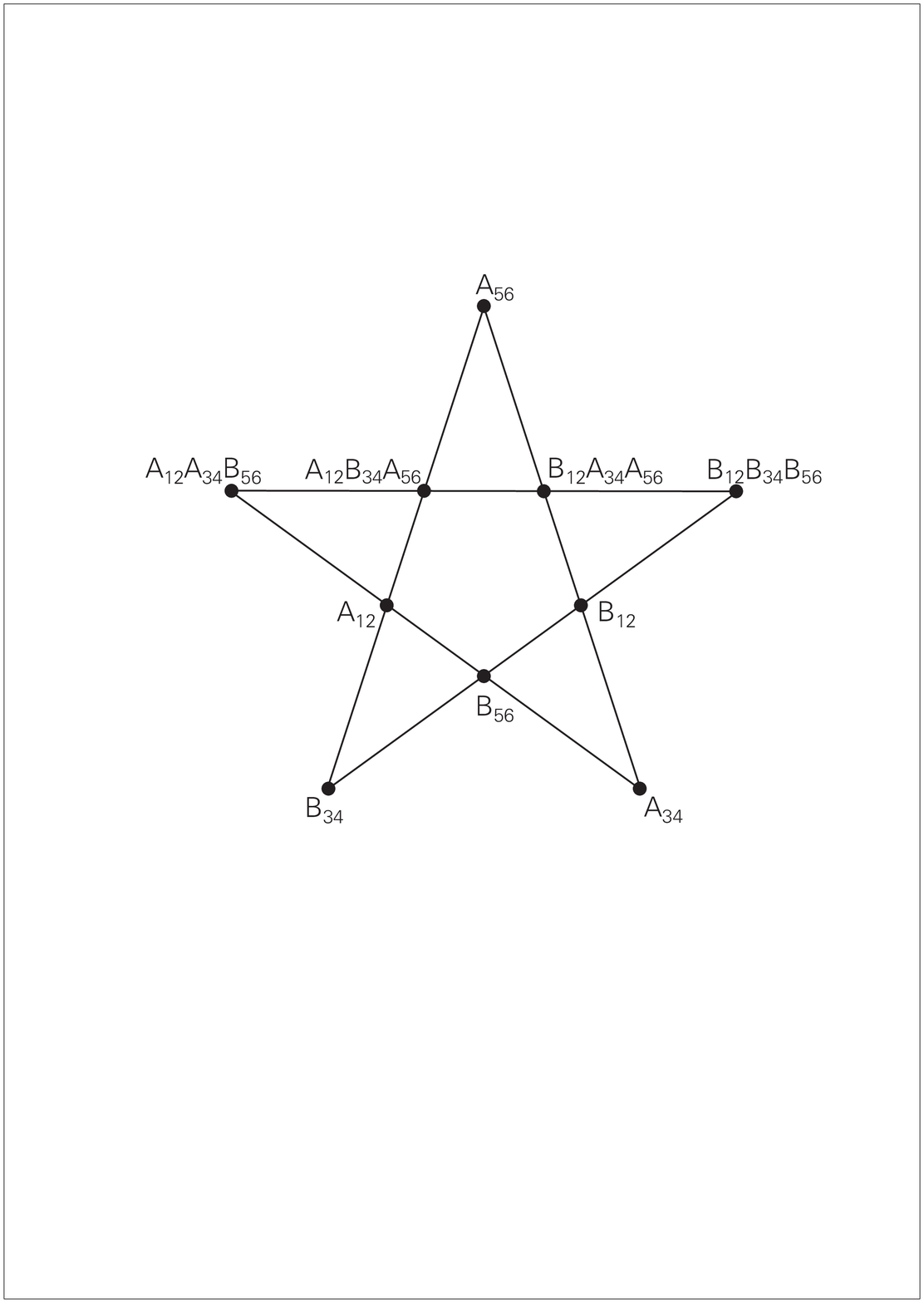}
\end{figure}
\noindent FIG. 1:
{\small Each dot represents an observable.
The ten observables provide a proof of
the Kochen-Specker theorem
in a Hilbert space of dimension sixty four.
The four observables  on each line are mutually compatible
and the product
of their results must be positive, except for the
horizontal line, where the product must be negative.}

\end{document}